\documentclass[aps,prc,reprint,showpacs,superscriptaddress,onecolumn,notitlepage,nofootinbib]{revtex4-2}
\usepackage{url,hyperref,lineno,microtype,subcaption}
\usepackage{graphicx}
\usepackage{dcolumn}
\usepackage{amsmath}
\usepackage{amssymb}
\usepackage{psfrag}
\usepackage{epsfig}
\usepackage{epsf}
\usepackage{float}
\newcommand{\nmax}{\ensuremath{N_{\rm max}}}
\newcommand{\hw}{\ensuremath{\hbar\omega}}
\begin{document}

\title[Uncertainties in ab initio nuclear structure calculations]{Uncertainties in ab initio nuclear structure calculations with chiral interactions}

\author{P.~Maris}
\affiliation{Dept. of Physics and Astronomy, Iowa State University, Ames, IA, USA}
\author{H.~Le}
\affiliation{Institut f\"ur Kernphysik, Institute for Advanced Simulation and J\"ulich Center for Hadron Physics, Forschungszentrum J\"ulich, J\"ulich, Germany}
\author{A.~Nogga}
\affiliation{Institut f\"ur Kernphysik, Institute for Advanced Simulation and J\"ulich Center for Hadron Physics, Forschungszentrum J\"ulich, J\"ulich, Germany}
\affiliation{Centre for Advanced Simulation and Analytics (CASA), Forschungszentrum J\"ulich, J\"ulich, Germany}
\author{R.~Roth}
\affiliation{Institut f\"ur Kernphysik, Technische Universit\"at Darmstadt, Darmstadt, Germany}
\author{J.P. Vary}
\affiliation{Dept. of Physics and Astronomy, Iowa State University, Ames, IA, USA}

\begin{abstract}

We present theoretical ground state energies and their uncertainties for $p$-shell nuclei obtained from chiral effective field theory internucleon interactions as a function of chiral order, fitted to two- and three-body data only.  We apply a Similary Renormalization Group transformation to improve the numerical convergence of the many-body calculations, and discuss both the numerical uncertainties arising from basis truncations and those from omitted induced many-body forces, as well as chiral truncation uncertainties.  With complete Next-to-Next-to-Leading (N$^2$LO) order two- and three-body interactions, we find significant overbinding for the ground states in the upper $p$-shell, but using higher-order two-body potentials, in combination with N$^2$LO three-body forces, our predictions agree with experiment throughout the $p$-shell to within our combined estimated uncertainties.  The uncertainties due to chiral order truncation are noticeably larger than the numerical uncertainties, but they are expected to become comparable to the numerical uncertainties at complete N$^3$LO.

\end{abstract}


\maketitle

\section{Introduction}

An atomic nucleus, consisting of $Z$ protons and $N$ neutrons, is a self-bound quantum many-body system with $A=N+Z$ strongly interacting nucleons.  The interactions between these nucleons are in principle governed by QCD -- but it is impractical to describe nuclei in terms of quarks and gluons, except for the very lightest systems.  Even a microscopic description of nuclei using realistic two-body (NN), three-body (3N) and possibly higher $n$-body interactions between point-like nucleons remains a formidable task, both in terms of high-performance computing, and in terms of determining realistic nuclear interactions in tractable terms.  In order to confront such a description with experimental data, one needs honest assessments of all uncertainties, both those arising from the numerical solution of a many-body problem, and those arising from a necessarily approximate theory of the effective interactions between nucleons.

Any ab initio theory of nuclei in terms of interacting nucleons requires a high-quality NN potential providing an accurate description of NN scattering data.  Highly accurate NN potentials have been in existence for several decades now, all incorporating one-pion exchange, and often inspired by one-boson-exchange (OBE) models, adjusted and augmented by phenomenological terms as necessary to fit the available NN data, such as the Argonne~\cite{Wiringa:1994wb}, (CD)-Bonn~\cite{Machleidt:1987hj,Machleidt:2000ge}, and Nijmegen~\cite{Stoks:1994wp} potentials.  Although there exist highly accurate NN potentials in terms of describing the NN phase-shift data, that at the same time accurately describe the spectra of light nuclei~\cite{Shirokov:2016ead}, most realistic NN potentials require more or less phenomenological 3N forces (3NFs)~\cite{Coon:1978gr,Pudliner:1997ck,Pieper:2001ap} for a good description of nuclei in the $p$-shell and beyond.  However, in order to quantify any uncertainties associated with the choice of the NN potential (and 3NFs), we need a more systematic method of arriving at the potential.

Chiral Effective Field Theory ($\chi$EFT) allows us to derive nuclear interactions in a systematic way, in terms of an expansion in the pion mass (or the relevant nucleon momentum) over the hadronic or breakdown scale \cite{Weinberg:1990rz,Epelbaum:2008ga,Machleidt:2011zz,Hammer:2019poc}; and in principle, it also allows for a quantification of the uncertainties arising from truncating this expansion.  However, this chiral expansion is by no means unique, and different choices for e.g.  the degrees of freedom to include in the $\chi$EFT can lead to very different $\chi$EFT interactions, with a different ordering of various types of diagrams, and indeed different orders at which higher $n$-body forces have to be included.  Furthermore, different choices on e.g. how to regulate the various expressions for loop integrals lead to different versions of the NN potentials (and 3NFs) at any given order, even if the ordering of the various diagrams is the same.  Each of these different versions of $\chi$EFT comes with its own parameters (Low-Energy Constants or LECs) that need to be fitted to data (or, eventually, calculated from e.g. lattice QCD), and with its own uncertainty quantification.  

Just like there are different ways to obtain (effective) nuclear interactions, there are different quantum many-body methods being used for ab initio nuclear structure calculations.  For up to four nucleons, one can use the Faddeev--Yakubovsky method (the 3- and 4-body reformulation of the Schr\"odinger equation that permits the incorporation of the appropriate boundary condition for 3- and 4-body systems that are asymptotically clustered), but this has not been extended to $A = 6$ or beyond.  Broadly speaking, the computational methods applicable to nuclei beyond $^4$He, fall into one of three categories: Quantum Monte Carlo simulations (both variational, VMC~\cite{Carlson:2014vla}, and Green's function, GFMC~\cite{Carlson:2014vla}), non-relativistic lattice simulations with nucleons (Nuclear Lattice Effective Field Theory, NLEFT~\cite{Laehde:2019bk,Lee:2020meg}), and Configuration Interaction (CI) methods (No-Core Shell Model (NCSM)~\cite{Barrett:2013nh}, Coupled-Cluster (CC)~\cite{Hagen:2013nca}, In-Medium Similarity Renormalization Group (IM-SRG)~\cite{Hergert:2015awm}), which are based on an expansion of the many-body wave-functions in terms of basis functions (configurations).  Each of these methods has their own uncertainties: Monte Carlo simulations are typically dominated by statistical uncertainties, though there is also a dependence on the variational wave function; lattice simulations have both statistical and systematic (lattice size and lattice spacing) uncertainties; and CI methods are generally dominated by systematic uncertainties due to the truncation of the many-body basis, though one can make use of statistical sampling of the many-body basis~\cite{OTSUKA2001319}.  Because each of these methods have different sources of uncertainties, and they are not always easy to identify and quantify, it is very valuable to use two or more of these many-body methods for the same nucleus, using the same interactions.

In this paper we use the NCSM to perform ab initio nuclear structure calculations for the ground state energies of nearly all stable $p$-shell nuclei (excluding mirror nuclei) from $A=4$ to $A=16$ using the $\chi$EFT interactions from Ref.~\cite{Reinert:2017usi}.  We perform a systematic set of order-by-order calculations in the chiral expansion to determine the uncertainties associated with the truncation of the chiral expansion; more details about the $\chi$EFT and how we estimate the truncation uncertainty can be found in Section~\ref{Sec:cEFT}.  In order to assess the numerical uncertainties in our NCSM calculations, we make a detailed comparison with Faddeev--Yakubovsky calculations for $^3$H and $^4$He using the same interactions; this is described in Section~\ref{Sec:NCSM}, together with details about the NCSM.  Our results for the binding energies of $p$-shell nuclei are presented in Section~\ref{Sec:results}.  Finally, we give some concluding remarks in Section~\ref{Sec:conclusion}.

\section{Nuclear interactions from chiral Effective Field Theory} \label{Sec:cEFT}

In recent years two different formulations of $\chi$EFT have emerged that are being used in ab initio nuclear structure calculations.  The most commonly used $\chi$EFT is based on only pions and nucleon degrees of freedom~\cite{Epelbaum:2008ga,Machleidt:2011zz}, for which the Leading Order (LO) and Next-to-Leading Order (NLO) terms consists of just two-body interactions; three-body interactions first appear at Next-to-Next-to-Leading order (N$^2$LO).  Alternatively, one can include $\Delta$ degrees of freedom into the EFT, in which case three-body interactions appear already at NLO~\cite{vanKolck:1994yi,Piarulli:2014bda}; see Refs.~\cite{Piarulli:2016vel,Piarulli:2017dwd} for nuclear structure calculations with these NN plus 3N interactions.
The reordering of contributions possibly speeds up the convergence of the chiral expansion. 

Here we use the formulation of $\chi$EFT based on only pion and nucleon degrees of freedom since high order potentials have already been developed for this approach.  This implies that we work with the conventional power-counting scheme, and with only NN potentials at LO and NLO, while 3N interactions arise at N$^2$LO.
Specifically, within the Low-Energy Nuclear Physics International Collaboration (LENPIC) we use the semilocal momentum-space (SMS) regulated NN potentials from Ref.~\cite{Reinert:2017usi}, which have been developed completely up through N$^4$LO; and the most accurate LENPIC-SMS NN potential, referred to as N$^4$LO$^+$, including some contributions from the 6th order in the chiral expansion.  The N$^4$LO$^+$ potential gives a near-perfect description of the mutually compatible neutron-proton and proton-proton scattering data below $E_{\hbox{\scriptsize lab}} = 300$~MeV with a $\chi^2_{\hbox{\scriptsize datum}}= 1.01$.  At the moment, the accompanying higher $n$-body forces have not yet been developed to the same chiral order.

Right now, consistent N$^2$LO 3NFs exist, implying that the regularization of the 3N interactions is consistent with that of the NN potential, that all relevant symmetries are respected, and that the same LEC values are used in the NN and 3N interactions.  The strength of the $2\pi$ exchange in the N$^2$LO 3NFs ($c_1$, $c_3$, and $c_4$) has been determined from $\pi$N scattering, see Table 1 of Ref.~\cite{Reinert:2017usi}.  (Note that, for the 3NF, these values need to be shifted as given in Eq.~(2.8) of Ref.~\cite{Bernard:2007sp}.)  We have not taken uncertainties of these $c_i$'s into account; this should be part of the N$^3$LO uncertainty estimate given below.  These 3NFs have already been used for nucleon-deuteron scattering~\cite{Epelbaum:2019zqc}, as well as select light nuclei~\cite{LENPIC:2022cyu,Maris:2020qne}.  Consistent N$^3$LO 3NFs are being developed and tested, and are expected to be available for use in many-body calculations soon; similarly, consistent electroweak operators are also under development.  

The 3NFs at N$^2$LO depend on two LECs, generally referred to as $c_D$ and $c_E$; these two LECs have been determined in Ref.~\cite{Maris:2020qne} by fitting the $^3$H binding energy (using the Faddeev approach), as well as the experimental proton-deuteron scattering data~\cite{Sekiguchi:2002sf} for the differential cross-section minimum at the proton beam energy of $E=70$~MeV.
Note that for the determination of $c_D$ and $c_E$ it is important to identify observables that (a) provide sufficiently independent constraints, i.e. are sensitive to the 3NFs and are sufficiently uncorrelated; (b) can be predicted accurately at N$^2$LO; and (c) are measured experimentally with sufficiently high accuracy. This can be achieved by e.g. incorporating properties of $^4$He (and other nuclei), in addition to $A=3$ observables, in the fitting of $c_D$ and $c_E$~\cite{Wesolowski:2021cni}.  However, here we prefer to only use $A=3$ data for the determination of $c_D$ and $c_E$ in order to obtain parameter-free predictions for $A > 3$, and to avoid interference of 4N (and higher-body) interactions at N$^3$LO and higher.  In Ref.~\cite{LENPIC:2018ewt}, it has been observed that the triton binding energy and the proton-deuteron scattering cross section minimum at 70 MeV are fulfilling these requirements.

Note that we keep all LECs in the NN potentials fixed at their values determined from NN scattering; and we do not propagate any uncertainties in these LECs through the many-body calculations.  Similarly, we have not explicitly propagated uncertainties in the LECs $c_D$ and $c_E$ for the 3NFs through the many-body calculations.  In Ref.~\cite{LENPIC:2018ewt} we did vary $c_D$ and $c_E$ while keeping the $^3$H binding energy fixed with the LENPIC Semilocal Coordinate Space interaction at N$^2$LO, and the resulting variation in the $^4$He and $^{12}$C binding energies, while not negligible and in opposite directions, stayed within the chiral truncation uncertainty estimate for a variation of $c_D$ between 6 and 8, the preferred range based on Nd scattering data for that interaction.
Furthermore, in Ref.~\cite{Carlsson:2015vda} it was shown that the uncertainties in many-body observables of $^4$He and $^{16}$O due to propagation of the uncertainties in determining the LECs at N$^2$LO are much smaller than the chiral truncation errors in those many-body observables at N$^2$LO.  We therefore assume here that any variation of the LECs of the NN and 3N interaction is an effect that is of higher order than N$^2$LO and thus those uncertainties are included in the uncertainty due to missing higher chiral orders.

\subsection{Chiral truncation uncertainty estimates}

Assuming that the chiral expansion of the nuclear interactions translates into a similar expansion for the physical observables, one expects that an observable $X$ follows a similar expansion pattern.  Consider therefore an observable $X$, and write it as
\begin{eqnarray}
    X & = & X^{(0)} + \Delta X^{(2)} + \Delta X^{(3)} + \dots \;,
\end{eqnarray}
where $X^{(0)}$ is the LO term, $\Delta X^{(2)} = X^{(2)} - X^{(0)}$ and $\Delta X^{(3)} = X^{(3)} - X^{(2)}$ are the NLO and N$^2$LO correction terms, respectively, and the dots represent higher-order corrections.  If this observable indeed follows the same expansion pattern as the nuclear interaction itself, then the correction terms $\Delta X^{(i)}$ behave like $Q^i$ for increasing $i$, where $Q = \max(p, M_\pi)/\Lambda_B$ is the chiral expansion parameter (typically the maximum of the relevant momentum $p$ and the pion mass $M_\pi$ over the breakdown scale $\Lambda_B$).  Note that there is no term linear in $Q$ in this expansion: the first correction, at NLO, is quadratic in the expansion parameter $Q$, at least for observables governed solely by the strong interaction.  For electroweak observables, the power-counting is different.

For the purpose of a Bayesian analysis, it is more convenient to rewrite this in terms of dimensionless expansion coefficient $c_i$, with the scale set by an overall reference value $X_{\hbox{\scriptsize ref}}$.  Thus we can rewrite the expansion for $X$ as
\begin{eqnarray} \label{eq:chiral_expansion}
    X & = & X_{\hbox{\scriptsize ref}} \left( c_0 + c_2 Q^2 + c_3 Q^3 + \dots \right) \;.
\end{eqnarray}
Now we can use Bayesian analysis on the coefficients $c_i$ to estimate the chiral truncation uncertainties.  Here we follow the Bayesian model of Ref.~\cite{Melendez:2019izc} for pointwise truncation errors with hyperparameters $\nu_0=1.5$ and $\tau_0=1.5$~\cite{Maris:2020qne}.   We apply this to the ground state energy of the $p$-shell nuclei, with the experimental value as our reference value $X_{\hbox{\scriptsize ref}}$.  Furthermore, we use an effective pion mass of $M_\pi^{\hbox{\scriptsize eff}} \approx 200$~MeV and a breakdown scale of $\Lambda_B \approx 650$~MeV~\cite{Epelbaum:2019zqc,Maris:2020qne}, and therefore a dimensionless expansion parameter $Q \approx 0.31$.  Note that in Ref.~\cite{LENPIC:2018lzt} it was observed that the average momentum of the nucleons inside a nucleus increases with $A$, and one might therefore have to increase $Q$ with $A$ as well; but up to $^{16}$O this average momentum remains below $200$~MeV so we use the same value for $Q$ throughout the $p$-shell.  Nevertheless, a Bayesian analysis of correlated uncertainties for ground states and excited states of a subset of $p$-shell nuclei does suggest a slightly larger value of $Q$ for the upper $p$-shell~\cite{LENPIC:2022cyu}.

Finally, although we use the LENPIC-SMS NN potentials from LO up to N$^4$LO$^+$, we only have the corresponding 3NFs at N$^2$LO.  We therefore perform our chiral truncation uncertainty analysis for the N$^2$LO through N$^4$LO$^+$ NN potentials, all in combination with the  N$^2$LO 3NFs, as if they were all N$^2$LO interactions; that is, we include only the coefficients $c_0$, $c_2$, and $c_3$ in Eq.~(\ref{eq:chiral_expansion}) (and again, there is no term linear in $Q$).

\section{No-Core Shell Model} \label{Sec:NCSM}

\subsection{Numerical method}

In the No-Core Shell Model (NCSM)~\cite{Barrett:2013nh}, the wavefunction $\Psi$ of a nucleus consisting of $Z$ protons and $N$ neutrons is expanded in a finite $A=Z+N$-body basis of Slater determinants $\Phi_k$ of single-particle wavefunctions $\phi_{nljm}(\vec{r})$
\begin{eqnarray}
  \Psi(\vec{r}_1, \ldots, \vec{r}_A) &=&
    \sum a_k \Phi_k(\vec{r}_1, \ldots, \vec{r}_A) \,.
\label{eq:basis_expansion}
\end{eqnarray}
With such an expansion, the many-body Schr\"odinger equation
\begin{eqnarray}
  {\bf\widehat{H}} \; \Psi(\vec{r}_1, \ldots, \vec{r}_A) &=&
     E \; \Psi(\vec{r}_1, \ldots, \vec{r}_A)
\label{eq:schrodinger}
\end{eqnarray}
becomes an eigenvalue problem
\begin{eqnarray}
  H_{ik} \; a_k &=& E \; a_i \,,
\label{eq:eigenvalueproblem}
\end{eqnarray}
for the coefficients $a_k$ of the expansion in Eq.~(\ref{eq:basis_expansion}).  The matrix $H_{ik}$ consists of matrix elements $\Phi_i \bf\widehat{H}$$\Phi_k$ (where integration over all spatial degrees of freedom is understood) of the many-body hamiltonian
\begin{eqnarray}
  {\bf\widehat{H}} &=& {\bf\widehat{T}}_{\hbox{\scriptsize rel}}
    + {\bf\widehat{V}}_{NN} + {\bf\widehat{V}}_{3N} + \ldots
\label{eq:hamiltonian}
\end{eqnarray}
consisting of the relative kinetic-energy operator, a two-body potential, and, in general three-body and higher $n$-body interaction terms.  If the interaction is limited to $n$-body terms, the matrix $H_{ik}$ for a nucleus with $A > n$ becomes a sparse matrix; in practice, the NCSM is generally applied with up to three-body interactions, and the corresponding Hamiltonian matrices are extremely sparse for $A \ge 6$.  For any finite basis expansion, the obtained eigenvalue $E$ gives a strict upper bound for the energy in the complete (though infinitely large) basis, at least for the lowest states of a given $Z$, $N$, and spin-parity quantum numbers $J^P$; and the corresponding eigenvector $\vec{a}$ gives an approximation to the $A$-body wavefunction $\Psi(\vec{r}_1, \ldots, \vec{r}_A)$.  As one increases the basis size, the obtained eigenvalues $E$ of the matrix $H_{ik}$  approach the exact eigenvalues for a given Hamiltonian ${\bf\widehat{H}}$.  

In the conventional NCSM one uses a harmonic oscillator (HO) basis for the single-particle wavefunctions $\phi_{nljm}(\vec{r})$, characterised by its scale parameter $\hbar\omega$.  One particular advantage of a HO basis is that one can treat the center-of-mass motion exactly: the Talmi--Moshinksy brackets~\cite{Talmi:1952,Moshinsky:1959qbh} can be used to convert between HO matrix elements in single-particle coordinates and relative plus center-of-mass coordinates; furthermore, with a many-body truncation on the total number of oscillator quanta in the many-body basis, the obtained wavefunctions factorize exactly into a center-of-mass wavefunction and a relative wavefunction~\cite{Lipkin:1958zz,Gloeckner:1974sst}.  The single-particle wavefunctions $\phi_{nljm}(\vec{r})$ are labelled by their radial quantum number $n$, orbital motion quantum number $l$, total single-particle spin $j=l\pm \frac{1}{2}$, and magnetic projection $m$ which satisfies $-j \le m \le j$.  In a HO basis, the combination $(2n+l)$ gives the number of HO quanta for each state; thus, in a HO basis with a truncation on $\sum_i(2n_i+l_i)$ over all $A$ nucleons, the factorization of the center-of-mass wavefunction is guaranteed.  We add a Lagrange multiplier acting on the center-of-mass coordinates of the many-body system to the Hamiltonian ${\bf\widehat{H}}$ to remove center-of-mass excited states from the low-lying spectrum~\cite{Lipkin:1958zz,Gloeckner:1974sst}; thus all low-lying states will have a $0s$ HO center-of-mass wavefunction.  Note that this does not alter the eigenvalues nor the eigenvectors for these states, it merely separates the center-of-mass excited states from the states with the lowest center-of-mass motion.

All NCSM calculations presented here were performed using the code Many-Fermion Dynamics--nuclear~physics~\cite{https://doi.org/10.1002/cpe.3129,SHAO2017,10.1007/978-3-319-46079-6_26}.  It solves the eigenvalue problem Eq.~(\ref{eq:eigenvalueproblem}) for the lowest eigenvalues, starting from two- and three-body matrix elements in a HO basis.  MFDn is a platform-independent Fortran~90 code using a hybrid MPI+OpenMP programming model.  The actual calculations have been performed on Theta at the Argonne Leadership Computing Facility (ALCF) and Cori at the National Energy Research Scientific Computing center (NERSC).  For each nucleus and interaction, we performed a series of calculations, using a range of different values of {\hw} and the truncation parameter \nmax, which is defined as the number of HO quanta above the minimal number of HO quanta in the many-body basis for that nucleus.  That is, an $\nmax=0$ calculation corresponds to a calculation in the lowest oscillator configuration.  Here we are only interested in the normal or natural parity states (the parity of the $\nmax=0$ space), and we increase \nmax\ in steps of 2 starting from $\nmax=0$ up to at least $\nmax=8$.  Some of the largest calculations for this study were for $^{14}$N and $^{15}$N at $\nmax=8$, both with dimensions of over one billion, and about $76 \times 10^{12}$ nonzero matrix elements, i.e. less than 1 in 10,000 matrix elements is nonzero with three-body interactions for these largest computations.

Of course, for two- and three-body systems it is more efficient and straightforward to work with wavefunctions in relative coordinates, rather than in single-particle wavefunctions.  However, beyond four nucleons, the necessary anti-symmetrization becomes increasingly cumbersome in relative coordinates, whereas the NCSM in single-particle coordinates is straightforward to implement for an arbitrarily large number of nucleons; however, the size of the matrix does grow dramatically with the number of nucleons.  Nevertheless, in recent years the NCSM has been implemented in Jacobi coordinates (J-NCSM)~\cite{Liebig:2015kwa} and applied to (hyper)nuclei with up to eight (hyper)nucleons~\cite{Le:2020zdu}.  The codes MFDn and J-NCSM have been benchmarked against each other, and generally agree to within 10 to 20 keV for $A=3$ and $4$, and to within about 30 keV for $A=6$, i.e to within 0.1\% of the obtained eigenvalues.  The differences of up to about 0.1\% have been attributed to differences in the implementations of transforming the three-body forces from their momentum-space expressions to HO matrix elements, including differences in the implementations of the Similarity Renormalization Group (SRG) transformations discussed next.

\subsection{Convergence and Similarity Renormalization Group evolution}

\begin{figure}[tb]
  \includegraphics[width=0.9\textwidth]{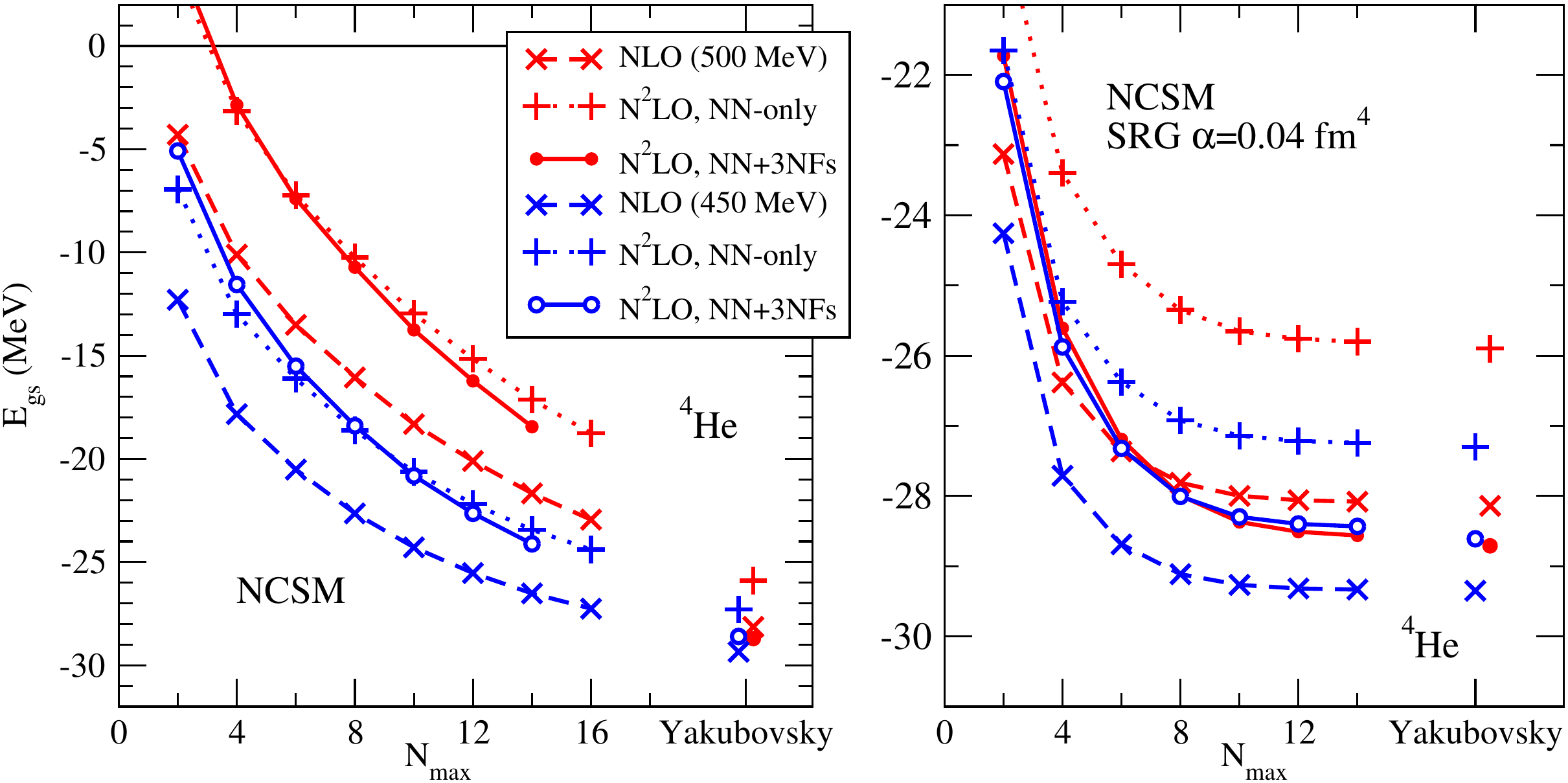}
  \caption{\label{Fig:res_4He_NCSM_Yak}
    Convergence of NCSM calculations for $^4$He without (left) and with (right) SRG evolution of the interactions.  
    Induced 3NFs are included in the calculations in the right-hand panel; for comparison results obtained with Yakubovsky calculations in momentum space are shown for the same interactions as well.}
\end{figure}
In the left panel of Fig.~\ref{Fig:res_4He_NCSM_Yak} we show the obtained ground state energy of $^4$He at NLO and N$^2$LO for two different values of the regulator $\Lambda$ as a function of \nmax\ at $\hw=24$~MeV; as illustration of the effect of the 3NFs, we also include results using only the NN potential at N$^2$LO, without the 3NFs (while at NLO, there are no 3NFs, so there is only the NN potential).  Even at $\nmax = 16$, the NCSM results are still several MeV above the corresponding Yakubovsky results, and far from being converged with \nmax; and for the upper half of the $p$-shell nuclei, for $A \ge 10$ we are restricted to $\nmax = 8$ in the presence of 3NFs due to computational limitations.   Clearly, we have to improve the numerical convergence while keeping the computational needs under control in order to obtain meaningful results for the ground state energies and other observables.  There are several methods to do so, which generally fall into four categories (and of course one can also use a combination of these techniques!)
\begin{itemize}
\item modify the underlying single-particle wavefunctions to improve the numerical convergence, e.g. start with a Hartree--Fock basis, and/or use natural orbitals~\cite{Constantinou:2016urz,Tichai:2018qge,Fasano:2021ahd};
\item modify the truncation scheme, e.g. select only the most important basis states at each step in \nmax\ (importance-truncated NCSM)~\cite{Roth:2009cw}, or use symmetries to reduce the number of basis states as \nmax\ increases (symmetry-adapted NCSM)~\cite{Dytrych:2007sv,McCoy:2018izw,Launey:2021sua};
\item reduce the 3N interaction to an effective NN interaction by normal-ordering the 3N interaction, which typically gains one step in \nmax\ in terms of computational needs~\cite{Hagen:2007ew,Roth:2011vt};
\item apply a unitary transformation on the Hamiltonian to improve the convergence at (relatively) small values of \nmax~\cite{Roth:2008km,Bogner:2009bt}.
\end{itemize}
Each of these methods has its advantages and drawbacks; and each of them is likely to obfuscate any uncertainty quantification of the numerical results; furthermore, with the first two methods listed above one might lose the exact factorization of the center-of-mass motion.  Here we choose to improve the numerical convergence in finite bases by applying a suitable SRG transformation on the Hamiltonian.

The SRG approach~\cite{Bogner:2009bt,Glazek:1993rc,Wegner:1994,Bogner:2007rx} provides a robust framework for consistently evolving (softening) the Hamiltonian, including three-body terms~\cite{Jurgenson:2009qs,Roth:2011ar,Roth:2013fqa,Binder:2013xaa}, as well as operators for other observables, by applying a unitary transformation on the operator(s) of interest.  This unitary transformation is formulated in terms of a flow equation
\begin{eqnarray}
  \frac{d}{d\alpha} {\bf\widehat{H}}_{\alpha} &=& [\eta_{\alpha},  {\bf\widehat{H}}_{\alpha}] \,,
  \label{eq:flow}
\end{eqnarray}
with a continuous flow parameter $\alpha$.  The physics of the SRG evolution is governed by the anti-hermitian generator $\eta_{\alpha}$.  A specific form widely used in nuclear physics~\cite{Bogner:2009bt} is given by
\begin{eqnarray}
  \eta_{\alpha} &=& m_N^2 [  {\bf\widehat{T}}_{\text{rel}},  {\bf\widehat{H}}_{\alpha}]\,,
\end{eqnarray}
where $m_N$ is the (average) nucleon mass and $ {\bf\widehat{T}}_{\text{rel}}$ is the relative kinetic-energy operator.  This generator drives the Hamiltonian towards a diagonal form in a basis of eigenstates of the intrinsic kinetic energy, i.e., towards a diagonal in momentum space.   The initial (or 'bare') Hamiltonian provides the initial condition at $\alpha=0$ for this flow equation; at NLO, this is just an NN-potential, but at N$^2$LO (and higher orders) it also includes the explicit 3NFs.
The width of the diagonal of the potential matrix elements in momentum space is proportional to  $\lambda_{\hbox{\tiny SRG}} = 1/\alpha^4$~\cite{Bogner:2007jb}.  For a typical value of $\alpha = 0.04$~fm$^4$, $\lambda_{\hbox{\tiny SRG}} \approx 2.24$~fm$^{-1}$, which can be considered as an effective cutoff in momentum space; lowering this cutoff improves the convergence of NCSM calculations.

Along with a decoupling of low-momentum and high-momentum components, this SRG induces many-body operators beyond the rank of the initial Hamiltonian.  In principle, all induced terms up to the $A$-body level should be retained to ensure that the transformation is a unitary transformation, such that the spectrum of the Hamiltonian is independent of the flow parameter $\alpha$.  In practice however, one has to truncate these many-body forces induced by the SRG evolution; here we follow the common practice of truncating the SRG evolution at the 3N level, omitting induced four-nucleon (and higher) induced interactions.  Of course, this violates unitarity, and therefore introduces a fictitious dependence on the SRG parameter $\alpha$ for $A \ge 4$ which we have to monitor, and include in our uncertainty budget. 
Unfortunately, it is as of yet unclear how to identify an expansion parameter that allows for an estimate of uncertainties due to missing higher-body induced interactions in $A \ge 4$ nuclei.

The flow equation for the three-nucleon system is solved numerically using a HO basis in Jacobi-coordinates~\cite{Roth:2013fqa} at a fixed HO basis parameter of $\hw=36$~MeV.  The intermediate sums in this three-body Jacobi basis are truncated at $N_{\max} = 40$ for channels with $J < 9/2$, $N_{\max}=38$ for $J=9/2$, and $N_{\max} = 36$ for all $J > 9/2$.  (Note that the flow equation at the two-nucleon level is solved numerically to a much higher numerical accuracy.)  The SRG evolution and transformations first from $\hw=36$~MeV to the desired $\hw$ value in the range from $14$ to $32$~MeV, and subsequently from Jacobi coordinates to single-particle coordinates, were all performed on a single multicore CPU node using an efficient OpenMP parallelized code.

In the right panel of Fig.~\ref{Fig:res_4He_NCSM_Yak} we show the ground state energy of $^4$He for the same initial interactions, and the same $\hw=24$~MeV, as in the left panel, but after first performing an SRG evolution of the Hamiltonian to $\alpha=0.04$~fm$^4$, and including induced 3N interactions, but omitting any induced 4N interactions.  Comparing these two panels it is immediately evident that the SRG evolution has indeed dramatically improved the convergence: after the SRG evolution to $\alpha=0.04$~fm$^4$, the obtained ground state energies at $\nmax=4$ are already closer to the Yakubovsky results than the ground states energies at $\nmax=16$ without any SRG evolution, and at $\nmax=14$ they appear to be almost converged and in agreement with the Yakubovsky results to within a fraction of an MeV.  Empirically, $\alpha$ in the range of $0.04 \leq  \alpha \leq 0.08$~fm$^4$ appears to be a good compromise between the convergence of the NCSM calculations and minimizing the contributions of the SRG-induced four- and higher-body forces.

\subsection{Extrapolating to the complete basis}

\begin{figure}[tb]
  \includegraphics[width=0.9\textwidth]{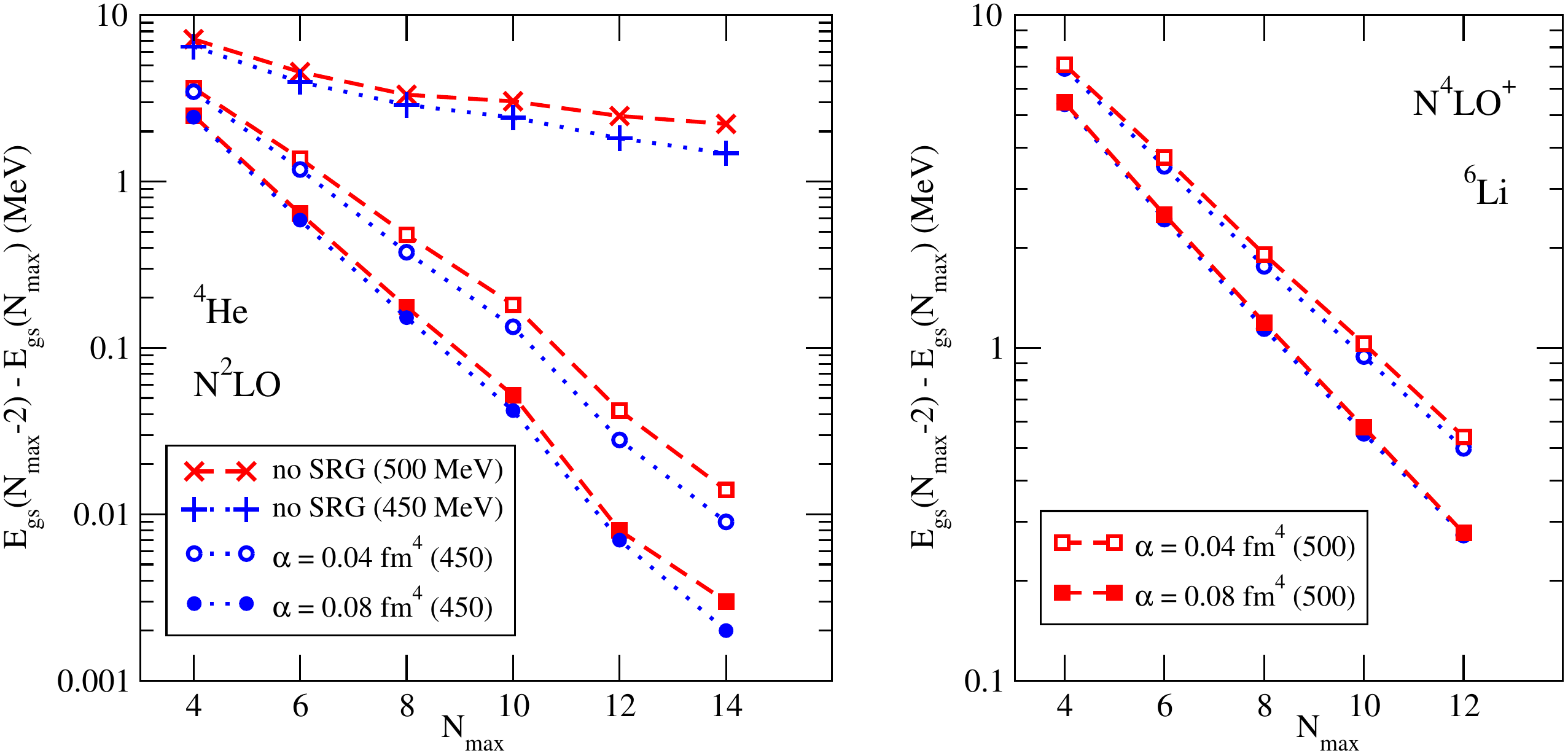}
  \caption{\label{Fig:res_NCSM_conv}
    Convergence of NCSM calculations for $^4$He (left) and $^6$Li (right): 
    Difference in obtained ground energies for successive \nmax\ values for different SRG evolution parameters $\alpha$.}
\end{figure}
Although the right-hand panel of Fig.~\ref{Fig:res_4He_NCSM_Yak} looks converged to well within 100 keV, it is not completely converged; furthermore, in the upper half of the $p$-shell we are limited to at most $\nmax = 8$, at which point the results are clearly not yet converged.  However, the approach to convergence appears to be smooth, and if we plot the difference between our results at successive \nmax\ values, at fixed \hw\ values near the variational minimum in the largest basis, see Fig.~\ref{Fig:res_NCSM_conv}, it is evident that these differences decrease almost exponentially with increasing \nmax.  Inspired by this behavior, we therefore use exponential extrapolation in \nmax\ at fixed \hw,
\begin{eqnarray} \label{eq:extrapolation}
  E^{\hw}(\nmax) &\approx & E^{\hw}_\infty + a\,{\rm e}^{(-b \nmax)} 
\end{eqnarray}
based on three consecutive values of \nmax\ at or slightly above the variational minimum in \hw\ to extract binding energies in the complete (but infinitely large) basis.  Indeed, such an empirical exponential has been widely used for a range of different interactions~\cite{Maris:2008ax,Jurgenson:2013yya,Maris:2019etr}, and appears to be reasonably reliable and accurate, at least for true bound states.  Furthermore, an exponential approach to convergence for the binding energy is also suggested by various analytic investigations into the asymptotic behavior~\cite{Coon:2012ab,Furnstahl:2012qg,More:2013rma,Wendt:2015nba}.
Insights into the approach to convergence allows one to improve the extrapolation~\cite{Forssen:2017wei}, but these analytic expressions generally depend on the underlying structure of the state.  Here, we restrict ourselves to the simple ansatz Eq.~(\ref{eq:extrapolation}) since it works well for all ground state energies considered, without the need to adapt the extrapolation to the specific structure of each nucleus.

Following Refs.~\cite{LENPIC:2022cyu,Maris:2020qne,LENPIC:2018lzt}, we take as our best estimate for $E_\infty$ in the complete basis the value of $E^{\hw}_\infty$ near the variational minimum in \hw\ for which $|E^{\hw}_\infty - E^{\hw}(\nmax)|$ is minimal.  Of course, this extrapolation is not exact, and will depend (slightly) on the \hw\ value; furthermore, we have to include an extrapolation uncertainty in our uncertainty budget.  Again, we resort to an empirical estimate of this uncertainty based on the variation with \hw\ and \nmax, and our estimate of the corresponding extrapolation uncertainty is the maximum of
\begin{itemize}
\item the difference in $E^{\hw}_\infty$ for two successive
  extrapolations using data for $(\nmax-6, \nmax-4, \nmax-2)$ and
  $(\nmax-4, \nmax-2, \nmax)$ respectively;
\item half the variation in $E^{\hw}_\infty$ over a 8~MeV interval in
  \hw\ around the variational minimum;
\item 20\% of $|E^{\hw}_\infty - E^{\hw}(\nmax)|$.
\end{itemize}
Note that this empirical uncertainty estimate is a conservative estimate, based on calculations with several different interactions~\cite{Maris:2019etr,Maris:2013poa}, and has been shown to give decreasing uncertainties with increasing \nmax, with the higher-\nmax\ results generally within the uncertainty estimates of the lower-\nmax\ results.  However, these uncertainty estimates cannot be interpreted statistically; for that one should use e.g. the Bayesian analysis of \cite{Gazda:2022fte}.

\subsection{Combined numerical uncertainties}

\begin{figure}[tb]
  \center\includegraphics[width=0.75\textwidth]{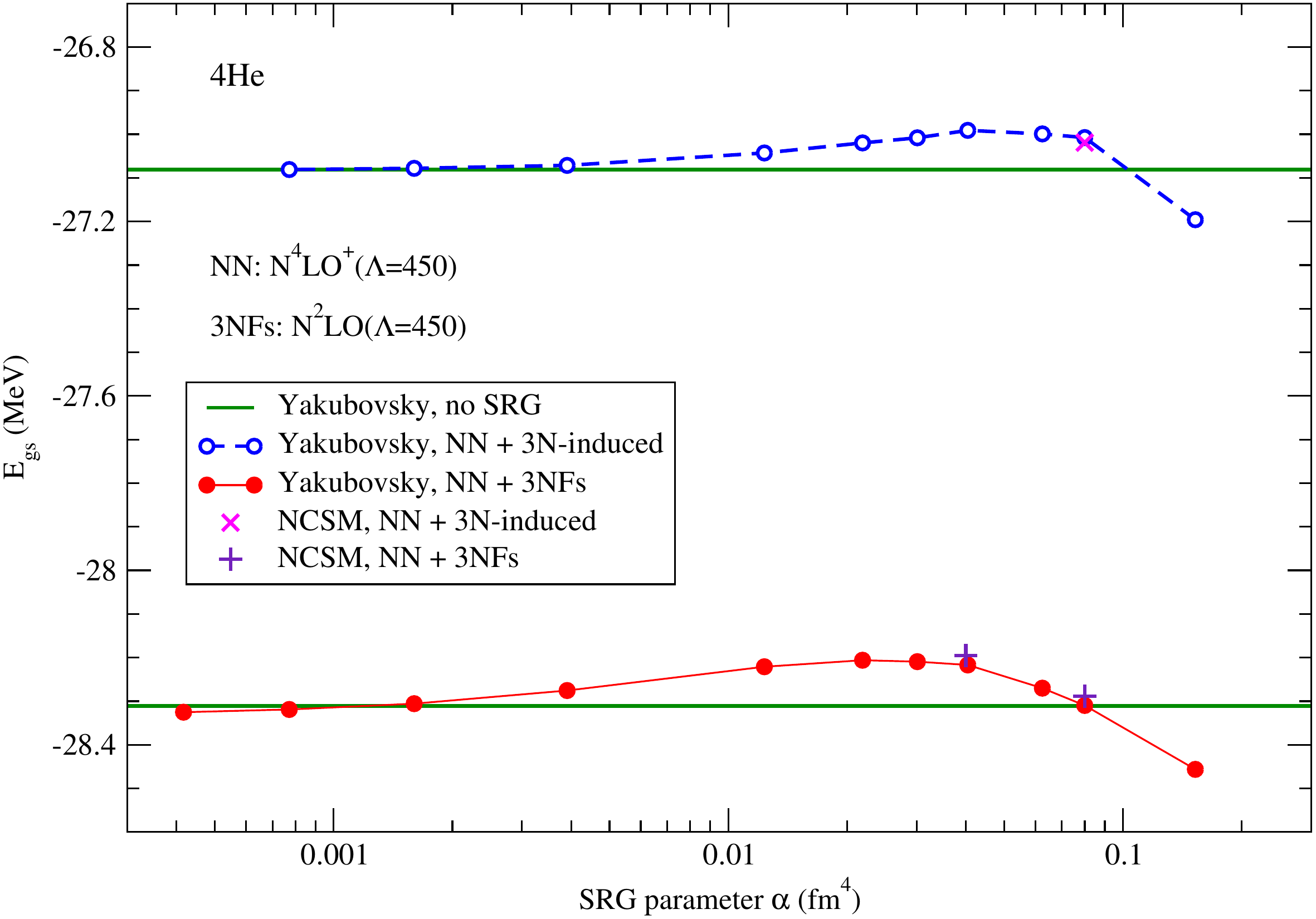}
  \caption{\label{Fig:res_4He_SRG}
    SRG depenence of the ground state energy of $^4$He, with the N$^4$LO$^+$ NN potentials plus the N$^2$LO 3NFs and $\Lambda=450$~MeV.
  }
\end{figure}

\begin{table}[tb]
  \begin{tabular}{c|D{.}{.}{6.8}|D{.}{.}{6.8}|D{.}{.}{6.8}|D{.}{.}{6.8}|D{.}{.}{6.8}}
    & \multicolumn{1}{c|}{NLO}
    & \multicolumn{1}{c|}{N$^2$LO}
    & \multicolumn{1}{c|}{N$^3$LO}
    & \multicolumn{1}{c|}{N$^4$LO}
    & \multicolumn{1}{c}{N$^4$LO$^+$}
    \\
    \hline\\[-9pt]
    $^3$H                & \multicolumn{5}{c}{ $\Lambda=450$~MeV } \\
    \hline\\[-9pt]
    Faddeev              & -8.515(.001)    & -8.483(.001)     & -8.483(.001)     & -8.483(.001)     & -8.483(.001)     \\
    $\alpha=0.04$~fm$^4$ & -8.54(.04)  & -8.51(.06)   & -8.50(.05)   & -8.51(.06)   & -8.50(.05)   \\
    $\alpha=0.08$~fm$^4$ & -8.517(.008) & -8.489(.017) & -8.483(.010) & -8.484(.010) & -8.488(.016) \\    
    \hline\\[-9pt]
    $^3$H                & \multicolumn{5}{c}{ $\Lambda=500$~MeV } \\
    \hline\\[-9pt]
    Faddeev              & -8.325(.001)    & -8.482(.001)      & -8.483(.001)      & -8.483(.001)      & -8.484(.001)  \\
    $\alpha=0.04$~fm$^4$ & -8.39(.10) & -8.52(.08)   & -8.51(.06)   & -8.51(.06)   & -8.51(.06)   \\
    $\alpha=0.08$~fm$^4$ & -8.327(.012)& -8.489(.019) & -8.485(.013) & -8.485(.011) & -8.491(.020) \\
    \hline\\[-9pt]
    $^4$He               & \multicolumn{5}{c}{ $\Lambda=450$~MeV } \\
    \hline\\[-9pt]
    Yakubovsky           & -29.36(.01)     & -28.61(.01)     & -28.35(.01)     & -28.29(.01)     & -28.31(.01)     \\
    $\alpha=0.04$~fm$^4$ & -29.339(.003) & -28.447(.004) & -28.284(.006) & -28.190(.004) & -28.195(.004) \\
    $\alpha=0.08$~fm$^4$ & -29.365(.001) & -28.527(.001) & -28.376(.002) & -28.285(.002) & -28.289(.002) \\
    $ |E(\alpha=0.04)-E(0)|$& 0.02    &   0.16     &   0.07     &   0.10     &   0.12     \\  
    $ |E(\alpha=0.08)-E(0)|$& 0.01    &   0.08     &   0.03     &   0.01     &   0.02     \\
    $ |E(0.04)-E(0.08)|$    & 0.01    &   0.08     &   0.04     &   0.09     &   0.10     \\
    \hline\\[-9pt]
    $^4$He               & \multicolumn{5}{c}{ $\Lambda=500$~MeV } \\
    \hline\\[-9pt]
    Yakubovsky           & -28.15(.01)      & -28.71(.01)      & -28.56(.01)      & -28.48(.01)      & -28.52(.01)      \\
    $\alpha=0.04$~fm$^4$ & -28.087(.003) & -28.585(.005) & -28.365(.003) & -28.236(.003) & -28.227(.003) \\
    $\alpha=0.08$~fm$^4$ & -28.122(.001) & -28.631(.003) & -28.447(.002) & -28.312(.002) & -28.301(.002) \\
    $ |E(\alpha=0.04)-E(0)|$& 0.06    &   0.13     &   0.20     &   0.24     &   0.29     \\  
    $ |E(\alpha=0.08)-E(0)|$& 0.03    &   0.08     &   0.11     &   0.17     &   0.22     \\
    $ |E(0.04)-E(0.08)|$    & 0.03    &   0.05     &   0.08     &   0.07     &   0.07     \\
    \hline
  \end{tabular}
  \caption{\label{Tab:SRG_dep_A3A4}
    SRG dependence of ground state energies in MeV for $A=3$ and $4$, compared to Faddeev--Yakubovsky calculations~\cite{LENPIC:2022cyu}.
    Explicit N$^2$LO 3NFs are included in the N$^2$LO through N$^4$LO$^+$ calculations.
    Quoted uncertainties are the estimated NCSM extrapolation uncertainties only.
  }
\end{table}
In Table~\ref{Tab:SRG_dep_A3A4} we give our extrapolated NCSM ground state energies for $^3$H and $^4$He, with our extrapolation uncertainty estimates, together with results obtained in momentum space with the Faddeev--Yakubovsky equations~\cite{LENPIC:2022cyu}.  For $^3$H the NCSM and Faddeev results agree very well, comfortably within the estimated extrapolation uncertainties of the NCSM calculations; and the results obtained with SRG $\alpha = 0.08$~fm$^4$ are more precise than those obtained with  $\alpha = 0.04$~fm$^4$, judging by their smaller uncertainties.  For the ground state energy of $^4$He we do see differences between the NCSM results and the Yakubovsky calculations, beyond the 10 keV uncertainty in the Yakubovsky calculations and the estimated NCSM uncertainties.  These differences can be attributed to the omitted induced 4NFs.  They are generally larger with the $\Lambda = 500$~MeV regulator than with the $\Lambda = 450$~MeV regulator, as one might expect, given that the $\Lambda = 500$~MeV interactions are converging slower than the $\Lambda = 450$~MeV interactions (see Fig.~\ref{Fig:res_4He_NCSM_Yak}); and it may be counter-intuitive that the effects of omitted 4NFs are larger for $\alpha = 0.04$~fm$^4$ than for $\alpha = 0.08$~fm$^4$, but this accidental, as can be seen from Fig.~\ref{Fig:res_4He_SRG}.  This figure also shows good agreement between the Yakubovsky and NCSM calculations.
 
\begin{table}[tb]
  \begin{tabular}{c|D{.}{.}{6.8}|D{.}{.}{6.8}|D{.}{.}{6.8}|D{.}{.}{6.8}|D{.}{.}{6.8}}
    & \multicolumn{1}{c}{NLO}
    & \multicolumn{1}{c}{N$^2$LO}
    & \multicolumn{1}{c}{N$^3$LO}
    & \multicolumn{1}{c}{N$^4$LO}
    & \multicolumn{1}{c}{N$^4$LO$^+$}
    \\
    \hline\\[-9pt]
    $^6$He & \multicolumn{5}{c}{ $\Lambda=450$~MeV } \\
    \hline
    $\alpha=0.04$~fm$^4$ & -28.73(.16) & -28.84(.17) & -28.16(.16) & -28.06(.16) & -28.11(.16) \\
    $\alpha=0.08$~fm$^4$ & -28.86(.14) & -29.05(.06)  & -28.39(.07)  & -28.28(.07)  & -28.33(.07)  \\
    $\Delta$             &   0.13     &   0.21     &   0.23     &   0.22     &   0.22     \\
    \hline\\[-9pt]
    $^6$He & \multicolumn{5}{c}{ $\Lambda=500$~MeV } \\
    \hline\\[-9pt]
    $\alpha=0.04$~fm$^4$ & -27.27(.15) & -29.08(.17) & -28.35(.17) & -28.19(.17) & -28.23(.16) \\
    $\alpha=0.08$~fm$^4$ & -27.39(.10) & -29.21(.06)  & -28.54(.06)  & -28.37(.06)  & -28.41(.07)  \\
    $\Delta$             &   0.12     &   0.13     &   0.23     &   0.22     &   0.22     \\
    \hline\\[-9pt]
    $^6$Li & \multicolumn{5}{c}{ $\Lambda=450$~MeV } \\
    \hline\\[-9pt]
    $\alpha=0.04$~fm$^4$ & -31.79(.11) & -31.85(.15) & -31.18(.14) & -31.07(.14) & -31.10(.14) \\
    $\alpha=0.08$~fm$^4$ & -31.93(.09)  & -32.04(.05)  & -31.41(.06)  & -31.28(.06)  & -31.32(.06)  \\
    $\Delta$             &   0.14     &   0.19     &   0.23     &   0.21     &   0.22     \\
    \hline\\[-9pt]
    $^6$Li & \multicolumn{5}{c}{ $\Lambda=500$~MeV } \\
    \hline\\[-9pt]
    $\alpha=0.04$~fm$^4$ & -30.33(.12) & -32.17(.16) & -31.42(.15) & -31.24(.15) & -31.26(.15) \\
    $\alpha=0.08$~fm$^4$ & -30.45(.06)  & -32.29(.05)  & -31.60(.06)  & -31.41(.05)  & -31.43(.05)  \\
    $\Delta$             &   0.12     &   0.12     &   0.18     &   0.17     &   0.17     \\
    \hline
  \end{tabular}
  \caption{\label{Tab:SRG_dep_A6}
    SRG dependence for $A=6$ ground state energies in MeV for SRG parameter $\alpha = 0.04$~fm$^4$ and $\alpha = 0.08$~fm$^4$, together with their difference.
    Quoted uncertainties are the estimated NCSM extrapolation uncertainties only.}
\end{table}
For $A \ge 6$ we do not have any calculations without SRG evolution for comparisons -- or rather, NCSM calculations for these interactions without SRG evolution are so far from convergence for $A \ge 6$ that they are not very useful for comparison.  However, we can gain insight in effects of omitted induced many-body forces by comparing results obtained with different values for the SRG parameter $\alpha$, see Tables~\ref{Tab:SRG_dep_A6} and \ref{Tab:SRG_dep_N2LO}.  Table~\ref{Tab:SRG_dep_A6} shows a differences of about $0.2$ MeV in the binding energies due to the two different SRG parameters $\alpha$, both for $^6$He and $^6$Li, and almost independent of the chiral order of the NN potential; though at NLO the difference is somewhat smaller, probably due to the lack of an explicit 3N interaction at NLO.  However, this difference of about $0.2$ MeV is the same order of magnitude as our estimated extrapolation uncertainties at $\alpha = 0.04$~fm$^4$, which prevents one from making firm conclusions.  

Note that the extrapolation uncertainties for $^6$He and $^6$Li at $\alpha=0.08$~fm$^4$ are a factor of two to three smaller than those obtained with $\alpha=0.04$~fm $^4$, clearly indicating the improved convergence as the interaction is further evolved with SRG.  The exception are the results for $A=6$ at NLO; this is most likely caused by the fact that the obtained binding energies at NLO are actually above threshold for $^6$He, and right around threshold for $^6$Li, as was already observed in Ref.~\cite{Maris:2020qne}.  Indeed, for states above threshold, the simple exponential extrapolation may not be very reliable since neglected continuum effects could be significant.

\begin{table}[tb]
  \begin{tabular}{c|D{.}{.}{6.8}|D{.}{.}{6.8}|D{.}{.}{6.8}|D{.}{.}{6.8}|D{.}{.}{6.8}|D{.}{.}{6.8}}
    & \multicolumn{1}{c|}{8Li}
    & \multicolumn{1}{c|}{10Be}
    & \multicolumn{1}{c|}{11B}
    & \multicolumn{1}{c|}{12C}
    & \multicolumn{1}{c|}{14C}
    & \multicolumn{1}{c}{15N}
    \\
    \hline\\[-9pt]
    $\alpha$ & \multicolumn{5}{c}{ $\Lambda=450$~MeV } \\
    \hline\\[-9pt]
    $0.04$ & -40.9(0.4) & -66.1(1.2) & -79.3(1.1) & -98.3(1.8) & -119.9(1.9) & -134.4(2.4) \\
    $0.08$ & -41.23(0.16)& -66.5(0.5)& -79.8(0.4) & -98.7(0.4) & -120.1(0.4) & -135.1(0.5) \\
    $\Delta$&  0.3      &   0.4      &   0.5      &   0.4      &    0.2      &    0.7      \\
    \hline\\[-9pt]
    $\alpha$ & \multicolumn{5}{c}{ $\Lambda=500$~MeV } \\
    \hline\\[-9pt]
    $0.04$ & -41.6(0.4) & -67.0(1.5) & -82.1(2.0) &-101.8(2.8) & -123.3(2.5) & -138.3(4.2) \\
    $0.08$ & -41.85(0.15)& -67.5(0.4)& -82.3(0.4) &-101.9(0.4) & -123.9(0.4) & -138.9(0.5) \\
    $\Delta$&  0.3      &    0.5     &   0.2      &   0.1      &    0.6      &    0.6    \\
    \hline    
  \end{tabular}
  \caption{\label{Tab:SRG_dep_N2LO}
    SRG dependence for select $6 < A < 16$ ground state energies in MeV at N$^2$LO for SRG parameter $\alpha = 0.04$~fm$^4$ and $\alpha = 0.08$~fm$^4$, together with their difference.
    Quoted uncertainties are the estimated NCSM extrapolation uncertainties only.}
\end{table}
In Table~\ref{Tab:SRG_dep_N2LO} we show the ground state energies for selected stable nuclei with $6 < A < 16$ using the N$^2$LO interaction (including 3NFs), SRG evolved to $\alpha = 0.04$~fm$^4$ and $\alpha = 0.08$~fm$^4$.  As is the case for $A=6$, the convergence improves with the SRG evolution: the extrapolation uncertainty estimates are a factor of 3 to 8 smaller at $\alpha = 0.08$~fm$^4$ than at $\alpha = 0.04$~fm$^4$.  This effect of improved convergence becomes more pronounced as $A$ increases, and is consistent with what we saw for $A=6$ in Table~\ref{Tab:SRG_dep_A6}.  Furthermore, at $\alpha = 0.04$~fm$^4$, starting from $A=10$, the $\Lambda=450$~MeV interaction converges noticeably better than the $\Lambda=500$~MeV interaction; in qualitative agreement with the picture for $^4$He (see Fig.~\ref{Fig:res_NCSM_conv}); however, at $\alpha = 0.08$~fm$^4$ this difference has washed away, and both regulators give a similar level of convergence for the ground state energies.

Somewhat surprisingly, the difference in ground state energies between the two SRG values, $\alpha = 0.04$~fm$^4$ and $\alpha = 0.08$~fm$^4$, remains almost constant, at around $0.5$~MeV, from $A=10$ to $A=15$, at least for the nuclei considered in Table~\ref{Tab:SRG_dep_N2LO}, with only a slight tendency to increase with $A$, and with similar tendencies for both regulator values.  Furthermore, this difference is similar to the estimated NCSM uncertainty at $\alpha = 0.08$~fm$^4$ (which also increases slowly with $A$), but a factor of 3 to 8 smaller than the NCSM uncertainty at $\alpha = 0.04$~fm$^4$, which makes it hard to draw a firm conclusion.  Nevertheless, based on these observations, we conclude that, for the calculations described here, it is realistic to include an SRG uncertainty that is equal to the NCSM extrapolation uncertainty estimate for stable $A=10$ to $A=16$ $p$-shell nuclei.

On the other hand, for $A=6$ and $^8$Li the difference in ground state energies between the two SRG values is noticeably larger than the NCSM extrapolation uncertainty for $\alpha = 0.08$~fm$^4$: for $^8$Li it is a factor of two larger; and for $A=6$ it is approximately a factor of three larger.  Note that this coincides with a larger \nmax\ value used for the lower half of the $p$-shell: for $A=6$ and $7$ we can perform our calculations up to $\nmax=12$ (and are in fact limited by the size of the input files with three-body HO matrix elements) and the calculations for $A=8$ and $9$ extend up to $\nmax=10$, whereas for $A \ge 10$ we are limited to $\nmax=8$.  Of course this upper limit in $\nmax$ also determines the level of numerical convergence that can be achieved, and hence the order of magnitude of the NCSM extrapolation uncertainty.  Again, based on these observations we estimate the SRG uncertainty in the binding energy to be about $0.2$~MeV for $A=6$ and $7$, and about $0.3$~MeV for $A=8$ and $9$.

\section{Ground state energies of p-shell nuclei} \label{Sec:results}

\begin{table}[b]
  \begin{tabular}{l|D{.}{.}{6.12}|D{.}{.}{6.12}|D{.}{.}{6.12}|D{.}{.}{6.12}}
    $V_{\rm NN}$
    & \multicolumn{1}{c|}{$^{4}$He$(0^+)$}
    & \multicolumn{1}{c|}{$^{6}$He$(0^+)$}
    & \multicolumn{1}{c|}{$^{6}$Li$(1^+)$}
    & \multicolumn{1}{c}{$^{7}$Li$(\frac{1}{2}^-)$}
    \\
    \hline \\[-7pt]
    & \multicolumn{4}{c}{$\Lambda = 450$~MeV} \\
    \hline \\[-7pt]
    LO         & -49.73(0.20)(-)   & -46.7^*(0.4)(-)    & -50.4^*(0.4)(-)   & -61.35^*(0.25)(-) \\
    NLO        & -29.37(0.20)(4.3) & -28.86^*(0.24)(3.9)& -31.93(0.22)(4.0) & -38.72(0.22)(4.9) \\
    N$^2$LO    & -28.53(0.20)(1.2) & -29.04(0.21)(1.0)  & -32.04(0.21)(1.1) & -39.39(0.21)(1.3) \\
    N$^3$LO    & -28.38(0.20)(1.2) & -28.39(0.21)(1.0)  & -31.41(0.21)(1.1) & -38.43(0.21)(1.3) \\
    N$^4$LO    & -28.29(0.20)(1.2) & -28.28^*(0.21)(1.0)& -31.28(0.21)(1.1) & -38.25(0.21)(1.3) \\
    N$^4$LO$^+$& -28.29(0.20)(1.2) & -28.33(0.21)(1.0)  & -31.32(0.21)(1.1) & -38.28(0.21)(1.3) \\
    \hline \\[-7pt]
    & \multicolumn{4}{c}{$\Lambda = 500$~MeV} \\
    \hline \\[-7pt]
    LO         & -51.17(0.20)(-)   & -47.6^*(0.5)(-)    & -51.1^*(0.4)(-)   & -62.1^*(0.3)(-)   \\
    NLO        & -28.12(0.20)(4.9) & -27.39^*(0.21)(4.3)  & -31.45(0.21)(4.2) & -36.82(0.23)(5.4) \\
    N$^2$LO    & -28.63(0.20)(1.3) & -29.21(0.21)(1.2)  & -32.29(0.20)(1.1) & -39.73(0.21)(1.5) \\
    N$^3$LO    & -28.45(0.20)(1.3) & -28.54(0.21)(1.2)  & -31.61(0.21)(1.1) & -38.72(0.21)(1.5) \\
    N$^4$LO    & -28.31(0.20)(1.3) & -28.37(0.21)(1.2)  & -31.41(0.21)(1.1) & -38.42(0.21)(1.5) \\
    N$^4$LO$^+$& -28.30(0.20)(1.3) & -28.41(0.21)(1.2)  & -31.43(0.21)(1.1) & -38.43(0.21)(1.5) \\
    \hline \\[-7pt]
    Expt.      & -28.30            & -29.27             &  -31.99           & -39.24       \\
    \hline
    \end{tabular}
  \caption{\label{Tab:Egs_A4A6A7}
    Ground state energies in MeV of $^4$He, $^6$He, $^6$Li, and $^7$Li, for LO through N$^4$LO$^+$ NN potentials, with 3NFs at N$^2$LO, for N$^2$LO through N$^4$LO$^+$, for $\Lambda=450$~MeV (top) and $\Lambda=500$~MeV (bottom).
    Both our estimated numerical uncertainties (first set of uncertainties) and chiral truncation uncertainty estimates (second set of uncertainties, not available for LO) are given.
    Entries with an $^*$ indicate energies above threshold, indicating a resonance, rather than a bound state.
  }
\end{table}
In Tables~\ref{Tab:Egs_A4A6A7}--\ref{Tab:Egs_A14A15A16} we present our results for the ground state energies of most stable $p$-shell nuclei, excluding mirror nuclei.  We also include $^8$Be, despite it being above the $2\alpha$ threshold.  All calculations were done in the NCSM approach, extrapolated to the (infinitely-large) complete basis, using NN (and 3N) potentials, SRG evolved to $\alpha=0.08$~fm$^4$, including induced 3N interactions, but omitting higher-body induced interactions.  The first set of uncertainties in these tables is our estimate of the combined numerical uncertainties; the second is our estimate of the chiral truncation uncertainty; both as described in the previous section.
The numerical uncertainty is estimated strictly based on the numerical convergence pattern and the SRG dependence, and cannot be interpreted statistically.  The chiral truncation uncertainty is based on a Bayesian model.  We give here the 68\% degree of belief (DoB) values.

The NCSM calculations for the nuclei presented in Table~\ref{Tab:Egs_A4A6A7} were performed up to $\nmax=14$ for $^4$He, and up to $\nmax=12$ for $A=6$ and $7$, which is generally sufficient to reach convergence for the ground state energies to within 0.1\% (or even better) for a given set of input HO two- and three-body matrix elements, thanks to the interaction being SRG evolved to $\alpha=0.08$~fm$^4$.  Therefore, the numerical uncertainties (the first set of quoted uncertainties in Table~\ref{Tab:Egs_A4A6A7}) are dominated by the uncertainties in the SRG evolution, which is mostly coming from the omission of induced 4-body forces (as well as higher-body forces for $A>4$), as well as from the numerical implementation of the SRG evolution and transformations from momentum space expressions to HO matrix elements. 

The exceptions are the $A=6$ and $7$ ground state energies at LO, because it turns out that at LO, these states are not bound, as indicated by the $^*$ in the tables: they are all above threshold for decays into $\alpha$ plus two neutrons, or plus a deuteron or a triton, respectively.  Hence the numerical convergence of the NCSM calculations is poor (at best it would converge to a quasi-bound state), and neither the extrapolation to the complete basis (nor its uncertainty estimate) is likely to be accurate, which is why the extrapolation uncertainties at LO are noticeably higher than at higher chiral orders.  
Since we only include the LO results to improve our estimate of the the chiral truncation uncertainties, an approximate bound state, or rather, resonance energy is sufficient for our purpose.  Similarly, $^6$He appears to be unbound at NLO; but again, for estimating the chiral uncertainty at N$^2$LO that is not a real problem. 

The estimated chiral truncation uncertainties are significantly larger than any of the numerical uncertainty estimates.  At NLO, these uncertainties are too large to draw any meaningful conclusions, but at N$^2$LO they are, as expected, more than a factor of three smaller.  Remember that we use N$^2$LO 3NFs in combination with the higher-order NN potentials, and we therefore apply the N$^2$LO power-counting rules for estimating the chiral uncertainties for these higher-order NN potentials.  It should therefore not be surprising that the obtained chiral uncertainty estimates are the same at these higher order as those with the N$^2$LO NN potential.  The central values however do change: all of the $A=4$, $6$, and $7$ nuclei become less bound when using NN potentials beyond N$^2$LO in combination with the N$^2$LO 3NFs.  This brings the ground state energy of $^4$He in closer agreement with experiment, whereas the ground state energies of $^6$He, $^6$Li, and $^7$Li are reasonably close to experiment with the N$^2$LO NN potential, and move away from their experimental values when including higher-order for the NN potential, to the point that $^6$He appears to be barely bound, or maybe even slightly unbound, with these higher-order NN potentials.  However, they all still agree with their corresponding experimental values, to within our combined numerical and chiral uncertainty estimates, and it is therefore too early to draw firm conclusions.

Finally, it is interesting to note that the estimated chiral truncation uncertainties are very similar for each of the four nuclei in Table~\ref{Tab:Egs_A4A6A7}.  This can be easily understood in terms of their structure: $^6$He, $^6$Li, and $^7$Li can be described as bound states of an $\alpha$ plus two neutrons, an $\alpha$ plus a deuteron, and an $\alpha$ plus a triton, respectively.  It is therefore not surprising that the chiral uncertainties of these states follow that of the $^4$He ground state energy (remember, the deuteron binding energy is fitted exactly, and the triton binding energies is fitted at N$^2$LO and up).  However, there are subtle but important details that can make a difference:  whereas the ground state energy of $^4$He changes only by about 150 to 200 keV going from the N$^2$LO to the N$^3$LO NN potential, the difference between these two potentials for the $A=6$ ground state energies is about 600 to 700 keV, and that for the $^7$Li ground state is about 1 MeV, for both regulators.

\begin{table}[tb]
  \begin{tabular}{l|D{.}{.}{6.12}|D{.}{.}{6.12}|D{.}{.}{6.12}|D{.}{.}{6.12}}
    $V_{\rm NN}$
    & \multicolumn{1}{c|}{$^{8}$He$(0^+)$}
    & \multicolumn{1}{c|}{$^{8}$Li$(2^+)$}
    & \multicolumn{1}{c|}{$^{8}$Be$(0^+)$}
    & \multicolumn{1}{c}{$^{9}$Li$(\frac{3}{2}^-)$}
    \\
    \hline \\[-7pt]
    & \multicolumn{4}{c}{$\Lambda = 450$~MeV} \\
    \hline \\[-7pt]
    LO         & -41.6^*(0.9)(-)   & -59.5^*(0.4)(-)   & -95.7^*(0.7)(-)    & -60.0^*(0.4)(-)   \\
    NLO        & -28.2^*(0.7)(3.0) & -39.44(0.36)(4.5) & -56.70^*(0.36)(8.3)& -41.55(0.45)(4.2) \\
    N$^2$LO    & -30.42(0.36)(0.9) & -41.23(0.34)(1.2) & -56.48^*(0.38)(2.2)& -45.14(0.34)(1.3) \\
    N$^3$LO    & -28.69(0.38)(0.8) & -39.62(0.34)(1.2) & -55.31^*(0.42)(2.2)& -42.27(0.37)(1.1) \\
    N$^4$LO    & -28.62(0.38)(0.8) & -39.45(0.34)(1.2) & -54.95^*(0.42)(2.2)& -42.11(0.36)(1.1) \\
    N$^4$LO$^+$& -28.75(0.38)(0.8) & -39.53(0.34)(1.2) & -54.98^*(0.42)(2.2)& -42.24(0.37)(1.1) \\
    \hline \\[-7pt]
    & \multicolumn{4}{c}{$\Lambda = 500$~MeV} \\
    \hline \\[-7pt]
    LO         & -41.6^*(1.0)(-)   & -59.6^*(0.4)(-)   & -97.7^*(0.7)(-)    & -59.8^*(0.5)(-)   \\
    NLO        & -26.3^*(0.6)(3.4) & -37.24(0.34)(4.9) & -53.77^*(0.36)(9.4)& -38.94(0.38)(4.7) \\
    N$^2$LO    & -30.92(0.34)(1.2) & -41.85(0.34)(1.5) & -56.96^*(0.37)(2.5)& -46.18(0.33)(1.8) \\
    N$^3$LO    & -29.06(0.36)(1.0) & -39.94(0.36)(1.4) & -55.70^*(0.38)(2.5)& -43.06(0.36)(1.4) \\
    N$^4$LO    & -28.91(0.35)(1.0) & -39.65(0.36)(1.4) & -55.10^*(0.38)(2.5)& -42.80(0.36)(1.4) \\
    N$^4$LO$^+$& -29.04(0.34)(1.0) & -39.72(0.36)(1.4) & -55.09^*(0.42)(2.5)& -42.91(0.36)(1.4) \\
    \hline \\[-7pt]
    Expt.      &  -31.41           &  -41.28           &  -56.50            & -45.32           \\
    \hline
  \end{tabular}
  \caption{\label{Tab:Egs_A8A9}
    Ground state energies in MeV of $A=8$ and $^9$Li, for LO through N$^4$LO$^+$ NN potentials, with 3NFs at N$^2$LO, for N$^2$LO through N$^4$LO$^+$, for $\Lambda=450$~MeV (top) and $\Lambda=500$~MeV (bottom).
    Both our estimated numerical (first) and chiral truncation (second) uncertainties are given;
    and an $^*$ indicates ground states with energies above threshold.
  }
\end{table}
In Table~\ref{Tab:Egs_A8A9} we give our results the ground state energies of $^8$He,  $^8$Li, $^8$Be, and $^9$Li.  The NCSM calculations for these nuclei were performed up to $\nmax=10$, making the extrapolation uncertainties for these nuclei somewhat larger than those in Table~\ref{Tab:Egs_A4A6A7}, and they become of the same order as the estimated numerical uncertainties coming from the SRG evolution.  This increased numerical uncertainty is reflected in the first set of error estimates in Table~\ref{Tab:Egs_A8A9}.  And just as for the $A=6$ and $7$ nuclei, at LO none of these nuclei are actually bound -- leading to larger extrapolation uncertainties.  Again, the main purpose for the LO calculations is to set the scale for the estimate of the chiral truncation uncertainties; and for that purpose, an approximate ground state energy is sufficient.  Furthermore, $^8$Be is unbound at all chiral orders considered here, in agreement with experiment.

As for the $A=4$, $6$, and $7$ nuclei, the estimated chiral truncation uncertainties for these $A=8$ and $9$ nuclei is significantly larger than the estimated numerical uncertainties.  Again, at NLO, these uncertainties are too large to draw any meaningful conclusions, but at N$^2$LO they are about a factor of three smaller.  The agreement with experiment is best with the N$^2$LO NN plus 3N interaction; moving to higher orders for the NN potential while retaining the N$^2$LO 3NFs leads to significant underbinding for $^8$He, $^8$Li, and $^9$Li, with the experimental values outside combined numerical and chiral uncertainty estimates for the N$^3$LO and higher NN potentials.  It will be very interesting to see whether or not consistent 3NFs at N$^3$LO can restore or improve on the level of agreement for these ground state energies obtained with N$^2$LO NN plus 3N interactions.

Beryllium-8 remains unbound according to our calculations, for all of these interactions, in qualitative agreement with experiment; and the extracted ground state energies may therefore be not as precise as for the other three nuclei in Table~\ref{Tab:Egs_A8A9}.  Still, given the combined numerical and chiral uncertainty estimates, our results for the $^8$Be ground state energy are in good agreement with experiment.  Furthermore it is interesting to note that the chiral uncertainty estimates for $^8$Be are approximately twice that of $^4$He, whereas the other three nuclei have chiral uncertainty estimates that are quite similar to those in Table~\ref{Tab:Egs_A4A6A7}.  This can be easily understood by realizing that $^8$Be is a loosely bound state, or rather, slightly unbound state, of two $\alpha$ particles, so the uncertainty is simply twice that of one $\alpha$ particle.  It may be more surprising that the chiral uncertainties of $^8$Li and $^9$Li, neither of which are $\alpha$-cluster states, are qualitatively similar to that of the $A=4$, $6$, and $7$; and it is also surprising that the estimated chiral uncertainties of $^8$He is smaller than that of any of the other $p$-shell nuclei.

\begin{table}[tb]
  \begin{tabular}{l|D{.}{.}{6.12}|D{.}{.}{6.12}|D{.}{.}{6.12}|D{.}{.}{6.12}}
    $V_{\rm NN}$
    & \multicolumn{1}{c|}{$^{9}$Be$(\frac{3}{2}^-)$}
    & \multicolumn{1}{c|}{$^{10}$Be$(0^+)$}
    & \multicolumn{1}{c|}{$^{10}$B$(3^+)$}
    & \multicolumn{1}{c}{$^{11}$B$(\frac{3}{2}^-)$}
    \\
    \hline \\[-7pt]
    & \multicolumn{4}{c}{$\Lambda = 450$~MeV} \\
    \hline \\[-7pt]
    LO         & -91.83^*(0.8)(-)   & -97.7^*(2.1)(-)& -92.8^*(2.3)(-) &-112.6(1.7)(-) \\
    NLO        & -56.97^*(0.36)(7.5)& -61.9(0.8)(7.8)& -61.1(0.8)(7.0) & -72.2(0.8)(8.9) \\
    N$^2$LO    & -58.82(0.37)(2.0)  & -66.5(0.7)(2.2)& -66.4(0.6)(2.1) & -79.8(0.6)(2.7) \\
    N$^3$LO    & -56.50^*(0.40)(2.0)& -62.4(0.8)(2.1)& -62.5(0.7)(1.9) & -73.8(0.8)(2.4) \\
    N$^4$LO    & -56.12^*(0.41)(2.0)& -62.0(0.8)(2.1)& -62.1(0.7)(1.9) & -73.4(0.8)(2.3) \\
    N$^4$LO$^+$& -56.18^*(0.41)(2.0)& -62.1(0.8)(2.1)& -62.2(0.8)(1.9) & -73.4(0.8)(2.3) \\
    \hline \\[-7pt]
    & \multicolumn{4}{c}{$\Lambda = 500$~MeV} \\
    \hline \\[-7pt]
    LO         & -93.00^*(0.7)(-)   & -98.1^*(2.4)(-)& -92.5^*(2.8)(-)  &-112.0^*(2.1)(-) \\
    NLO        & -53.68^*(0.37)(8.4)& -57.9(0.8)(8.6)& -57.0^*(0.7)(7.7)&-67.2(0.8)(9.7) \\
    N$^2$LO    & -59.73(0.36)(2.4)  & -67.5(0.6)(2.8)& -68.4(0.6)(2.8)  & -82.3(0.6)(3.6) \\
    N$^3$LO    & -57.20(0.40)(2.3)  & -63.6(0.8)(2.5)& -64.1(0.7)(2.4)  & -75.8(0.8)(3.0) \\
    N$^4$LO    & -56.59^*(0.40)(2.3)& -62.9(0.8)(2.4)& -63.4(0.8)(2.3)  & -74.7(0.7)(2.9) \\
    N$^4$LO$^+$& -56.61(0.40)(2.3)  & -63.0(0.8)(2.4)& -63.4(0.8)(2.3)  & -74.6(0.8)(2.9) \\
    \hline \\[-7pt]
    Expt.      &  -58.16            &  -64.98        &  -64.75          &  -76.21     \\
    \hline
  \end{tabular}
  \caption{\label{Tab:Egs_A9A10A11}
    Ground state energies in MeV of $^{9}$Be, $^{10}$Be, $^{10}$B, and $^{11}$B, for LO through N$^4$LO$^+$ NN potentials, with 3NFs at N$^2$LO, for N$^2$LO through N$^4$LO$^+$, for $\Lambda=450$~MeV (top) and $\Lambda=500$~MeV (bottom).
    Both our estimated numerical (first) and chiral truncation (second) uncertainties are given;
    and an $^*$ indicates ground states with energies above threshold.
  }
\end{table}
Moving to the middle of the $p$-shell, in Table~\ref{Tab:Egs_A9A10A11} we have our results for the ground state energies of $^9$He,  $^{10}$Be, $^{10}$B, and $^{11}$B.  Starting from $A=10$, the NCSM calculations are limited to $\nmax=8$, and therefore the extrapolation uncertainties become a significant factor in the uncertainty budget.  Nevertheless, qualitatively, the overall picture remains the same: at LO all nuclei are unbound, but at NLO and beyond, they are generally bound, with the exception of $^9$Be.  Also, the estimated chiral truncation uncertainties for these nuclei remains significantly larger than the estimated numerical uncertainties; at NLO, these uncertainties are too large to draw any meaningful conclusions, but at N$^2$LO and beyond they are about a factor of three smaller.  Here, we also start to see significant differences between the chiral uncertainties with the N$^2$LO NN plus 3N interaction, vs. using an NN potential at N$^3$LO or higher in combination with the N$^2$LO 3NFs (and remember, we are using the N$^2$LO counting rules for all these calculations with higher-order NN potential) -- the effect of the higher-order NN potentials is becoming more pronounced with increasing $A$, and more so with $\Lambda = 500$~MeV than with $\Lambda=450$~MeV.  (This trend is already noticeable for e.g. $^9$Li, see Table~\ref{Tab:Egs_A8A9}.)

Around $A=10$, the agreement with experiment is no longer uniformly better with the N$^2$LO NN plus 3N interaction than with the higher-order NN potentials.  The N$^2$LO NN plus 3N interaction is the only combination for which $^9$Be is truly bound with respect to two $\alpha$ particles plus a neutron, for both regulator values, whereas with the higher-order NN potentials $^9$Be becomes unbound or right at threshold in contrast with experiment where it is bound by about $1.6$~MeV.  Within the combined uncertainty estimates however, it is still in agreement with experiment for all of these interactions.  For the $A=10$ ground state energies the situation is different: at $\Lambda=450$~MeV, the N$^2$LO NN plus 3N interaction gives slightly better agreement with experiment than the higher-order NN potentials, but at $\Lambda=500$~MeV, it is the N$^3$LO NN plus N$^2$LO 3N interaction that gives the best agreement with experiment.  In fact, the $^{10}$B ground state energy with the N$^2$LO NN plus 3N potential is just outside the combined uncertainty estimates.  Furthermore, the N$^2$LO NN potential plus 3NFs give a ground state energy for $^{11}$B that is just outside the combined uncertainty estimates with both the $\Lambda=450$~MeV and $\Lambda=500$~MeV regulators.

\begin{table}[tb]
  \begin{tabular}{l|D{.}{.}{6.12}|D{.}{.}{6.12}|D{.}{.}{6.12}|D{.}{.}{6.12}}
    $V_{\rm NN}$
    & \multicolumn{1}{c|}{$^{12}$B$(1^+)$}
    & \multicolumn{1}{c|}{$^{12}$C$(0^+)$}
    & \multicolumn{1}{c|}{$^{13}$B$(\frac{3}{2}^-)$}
    & \multicolumn{1}{c}{$^{13}$C$(\frac{1}{2}^-)$}
    \\
    \hline \\[-7pt]
    & \multicolumn{4}{c}{$\Lambda = 450$~MeV} \\
    \hline \\[-7pt]
    LO         & -113.7(1.8)(-)   & -145.0^*(1.3)(-) & -120.8^*(1.7)(-) & -146.4^*(1.1)(-) \\
    NLO        &  -76.0(1.0)(8.4) &  -89.7(0.7)(12.0)&  -82.5(1.3)(8.6) &  -94.3(0.7)(11.4)\\
    N$^2$LO    &  -84.8(0.6)(2.7) &  -98.7(0.6)(3.5) &  -93.2(0.8)(2.9) & -108.3(0.6)(3.9) \\
    N$^3$LO    &  -77.3(0.8)(2.2) &  -90.6(0.8)(3.2) &  -83.2(1.0)(2.3) &  -96.7(0.8)(3.0) \\
    N$^4$LO    &  -76.8(0.8)(2.2) &  -89.9(0.8)(3.2) &  -82.6(1.0)(2.3) &  -96.2(0.8)(3.0) \\
    N$^4$LO$^+$&  -77.0(1.0)(2.2) &  -90.0(1.0)(3.2) &  -82.7(1.0)(2.3) &  -96.3(0.8)(3.0) \\
    \hline \\[-7pt]
    & \multicolumn{4}{c}{$\Lambda = 500$~MeV} \\
    \hline \\[-7pt]
    LO         & -111.7^*(2.3)(-) & -144.6^*(1.8)(-) & -117.4^*(2.1)(-) & -143.8^*(1.6)(-) \\
    NLO        &  -70.4(0.8)(9.1) &  -83.3^*(0.7)(13.1)&-76.1(1.1)(9.1) &  -87.0(0.7)(12.3)\\
    N$^2$LO    &  -87.5(0.6)(3.8) & -101.8(0.6)(4.7) &  -95.8(1.0)(4.2) & -112.2(0.6)(5.4) \\
    N$^3$LO    &  -79.5(0.8)(2.9) &  -92.7(0.8)(3.8) &  -85.5(1.0)(2.9) &  -99.8(0.7)(3.9) \\
    N$^4$LO    &  -78.8(0.8)(2.8) &  -91.6(0.8)(3.7) &  -84.6(1.0)(2.8) &  -98.8(0.7)(3.8) \\
    N$^4$LO$^+$&  -78.8(1.0)(2.9) &  -91.5(0.8)(3.7) &  -84.6(1.0)(2.8) &  -98.7(0.7)(3.8) \\
    \hline \\[-7pt]
    Expt.      &  -79.58          &  -92.16          &  -84.45          &  -97.11          \\
    \hline
  \end{tabular}
  \caption{\label{Tab:Egs_A12A13}
    Ground state energies in MeV of $A=12$ and $A=13$ nuclei, for LO through N$^4$LO$^+$ NN potentials, with 3NFs at N$^2$LO, for N$^2$LO through N$^4$LO$^+$, for $\Lambda=450$~MeV (top) and $\Lambda=500$~MeV (bottom).
    Both our estimated numerical (first) and chiral truncation (second) uncertainties are given;
    and an $^*$ indicates ground states with energies above threshold.
    }
\end{table}
%
\begin{table}[b]
  \begin{tabular}{l|D{.}{.}{6.12}|D{.}{.}{6.12}|D{.}{.}{6.12}|D{.}{.}{6.12}}
    $V_{\rm NN}$
    & \multicolumn{1}{c|}{$^{14}$C$(0^+)$}
    & \multicolumn{1}{c|}{$^{14}$N$(1^+)$}
    & \multicolumn{1}{c|}{$^{15}$N$(\frac{1}{2}^-)$}
    & \multicolumn{1}{c}{$^{16}$O$(0^+)$}
    \\
    \hline \\[-7pt]
    & \multicolumn{4}{c}{$\Lambda = 450$~MeV} \\
    \hline \\[-7pt]
    LO         & -160.2(0.7)(-)   & -160.1(0.8)(-)   & -182.1(0.6)(-)   & -218.3(0.3)(-)   \\
    NLO        & -104.9(0.8)(12.2)& -104.2(0.6)(12.3)& -119.4(0.8)(13.8)& -135.1(0.8)(17.9)\\
    N$^2$LO    & -120.1(0.6)(4.1) & -121.4(0.6)(4.4) & -135.1(0.5)(4.5) & -149.1(1.0)(5.3) \\
    N$^3$LO    & -106.3(0.8)(3.2) & -106.9(0.7)(3.3) & -118.8(0.8)(3.6) & -131.7(1.3)(4.8) \\
    N$^4$LO    & -105.5(0.8)(3.2) & -106.2(0.8)(3.3) & -117.8(0.8)(3.6) & -130.2(1.3)(4.8) \\
    N$^4$LO$^+$& -105.7(0.8)(3.2) & -106.4(0.8)(3.3) & -118.0(0.8)(3.6) & -130.4(1.3)(4.8) \\
    \hline \\[-7pt]
    & \multicolumn{4}{c}{$\Lambda = 500$~MeV} \\
    \hline \\[-7pt]
    LO         & -156.1(1.1)(-)   & -155.4(1.3)(-)   & -175.5(0.8)(-)   & -209.2(0.6)(-)   \\
    NLO        &  -96.2(0.7)(13.0)&  -95.5(0.6)(13.0)& -109.3(1.0)(14.4)& -123.5(1.0)(18.3)\\
    N$^2$LO    & -123.9(0.6)(5.9) & -125.6(0.6)(6.2) & -138.9(0.7)(6.3) & -153.2(1.4)(7.0) \\
    N$^3$LO    & -109.8(0.8)(4.1) & -110.8(0.8)(4.3) & -122.5(1.0)(4.4) & -135.4(1.4)(5.2) \\
    N$^4$LO    & -108.5(0.8)(4.0) & -109.5(0.8)(4.2) & -120.8(1.0)(4.3) & -133.0(1.7)(5.1) \\
    N$^4$LO$^+$& -108.5(0.8)(4.0) & -109.4(0.8)(4.2) & -120.7(1.0)(4.3) & -132.8(1.8)(5.1) \\
    \hline \\[-7pt]
    Expt.      & -105.28      	  & -104.66   	     & -115.49	      	& -127.62     	   \\
    \hline
  \end{tabular}
  \caption{\label{Tab:Egs_A14A15A16}
    Ground state energies in MeV of $^{14}$C, $^{14}$N, $^{15}$N, and $^{16}$O, for LO through N$^4$LO$^+$ NN potentials, with 3NFs at N$^2$LO, for N$^2$LO through N$^4$LO$^+$, for $\Lambda=450$~MeV (top) and $\Lambda=500$~MeV (bottom).
    Both our estimated numerical (first) and chiral truncation (second) uncertainties are given.
    }
\end{table}
This trend becomes more pronounced for $A \ge 12$, see Tables~\ref{Tab:Egs_A12A13} and \ref{Tab:Egs_A14A15A16}.  At LO, the ground states are still unbound, except for $^{12}$B at $\Lambda=450$~MeV; at NLO they are bound and in agreement with experiment, given the (granted, rather large) uncertainty estimates; and at N$^2$LO they are all significantly overbound, with the experimental values outside the combined numerical and chiral uncertainty estimates.  Increasing the chiral order of the NN potential improves the agreement with experiment again: for $A=12$ and $13$ our results with the N$^3$LO and higher NN potentials, in combination with the N$^2$LO 3NFs, agree with experiment, well within our uncertainty estimates, with both the $\Lambda=450$~MeV and $\Lambda=500$~MeV regulators.  For $A=14$ this is also the case with $\Lambda=450$~MeV, but $\Lambda=500$~MeV leads to modest overbinding, though still within our uncertainty estimates.  Also for $^{15}$N and $^{16}$O the ground state energies agree with experiment with $\Lambda=450$~MeV, but with $\Lambda=500$~MeV there is significant overbinding, with the experimental values just at the edge of our uncertainty intervals.

\begin{figure}
  \includegraphics[width=0.85\textwidth]{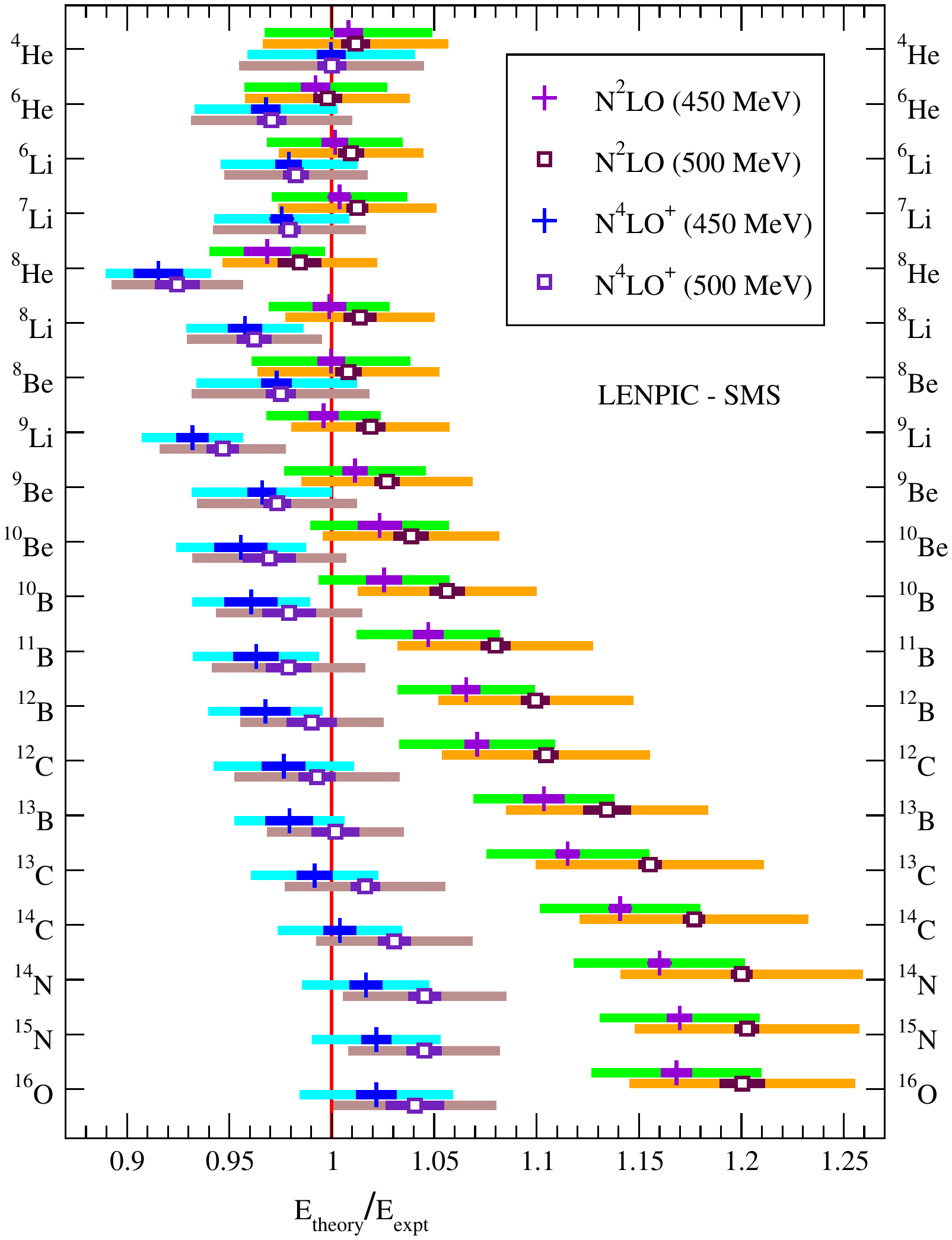}
  \caption{\label{Fig:res_Egs_ratio}
    Comparison of ground state energies of $p$-shell nuclei between chiral EFT calculations at N$^2$LO and N$^4$LO$^+$, each for two values of the regulator $\Lambda$, and experiment.  Both numerical uncertainty estimates (dark colored) and chiral truncation uncertainties (light colored, corresponding to 68\% DoB) are shown.
    }
\end{figure}
We have visually summarized our findings in Fig.~\ref{Fig:res_Egs_ratio}, which clearly shows that with the N$^2$LO NN plus 3N interaction one finds good agreement with experiment for the ground state energies of nuclei up to about $A=9$, but significant overbinding starting from about $A=11$, more than the estimated uncertainties for $A=13$ and beyond.  On the other hand, using higher-order NN potentials, in combination with  N$^2$LO 3NFs, reduces this overbinding in the upper half of the $p$-shell, while maintaining reasonable agreement, taking into account both numerical and chiral truncation uncertainties, in the lower half of the $p$-shell, with only a few exceptions, out of the 20 nuclei considered here.  

Clearly, for $A > 12$ our uncertainty estimates for the N$^2$LO NN plus 3N interaction are noticeably smaller than the deviation from both the experimental data and from the calculations with higher-order NN potentials.  We speculate that this may be caused by the N$^2$LO fit to the NN scattering data not being sufficiently accurate, and that discrepancies between the N$^2$LO fit and NN data should be taken into account as uncertainties in the LECs, whereas at higher orders in the NN potential, the NN scattering data are described much more accurately, and this is therefore not necessary.  
(Note that the N$^2$LO NN potential was fitted only up to $E_{\hbox{\scriptsize{lab}}} = 125$~MeV, whereas the N$^4$LO$^+$ potential was fitted to 260 MeV in Ref.~\cite{Reinert:2017usi}.)
Of course, one should then also incorporate the uncertainties in the 3NFs, $c_D$ and $c_E$~\cite{Wesolowski:2021cni}, and propagate all these uncertainties through the many-body bound state calculations~\cite{Carlsson:2015vda,Gazda:2022fte}.

Another possible explanation could be that NN (and 3N) systems cannot sufficiently constrain the LECs -- in which case one necessarily has to include properties of $A \ge 4$ nuclei for fitting some (or even all) of the LECs.  Indeed, impressive progress has been made in recent years along this way, extending ab initio calculations all the way to $^{208}$Pb~\cite{Hu:2021trw}, but one loses some of the predictive power of $\chi$EFT by incorporating select many-body observables in the fitting procedures, and the results will depend on exactly which observables are included in the fitting.  Yet another cause could be that the actual expansion parameter increases with $A$, as suggested in Ref.~\cite{LENPIC:2018lzt}. Calculations with consistent 3NFs at N$^3$LO, propagation of the uncertainties in the LECs through the many-body calculations, and Bayesian inference for both the chiral truncation uncertainties and the numerical uncertainties should help to resolve this issue.

Besides this general trend of increasing deviations with increasing $A$ at N$^2$LO, $^8$He and $^9$Li
clearly stand out among the N$^4$LO$^+$ results in Fig.~\ref{Fig:res_Egs_ratio}; and also our predictions for $^8$Li do not agree, to within their estimated uncertainties, with experiment.  Interestingly, $^8$He and $^9$Li are two of the most neutron-rich nuclei, out of the 20 nuclei shown in  Fig.~\ref{Fig:res_Egs_ratio}, with $N-Z=4$ and $3$, respectively; and also $^8$Li is a neutron-rich nucleus.  This could be an indication of some deficiencies in the neutron-neutron (or three-neutron) part of the interactions.  Unfortunately, there are no accurate neutron-neutron data, let alone three-neutron data, to constrain the LECs; the LECs of the interactions were all fitted to 2- and 3-body data involving at least one proton.

\section{Concluding remarks and outlook} \label{Sec:conclusion}

We have performed systematic calculations for the binding energies of $p$-shell nuclei using LENPIC-SMS $\chi$EFT NN and 3N interactions complete up through N$^2$LO, and with NN potentials up to N$^4$LO$^+$ in combination with N$^2$LO 3NFs.  We have made a careful analysis of all sources of uncertainties, and incorporated our best estimates of these uncertainties in our comprehensive tables with order-by-order results and in Fig.~\ref{Fig:res_Egs_ratio}.  Note that all LECs in the $\chi$EFT had been fitted to $A=2$ and $A=3$ data prior to these many-body calculations, and the obtained binding energies are therefore parameter-free predictions.  Although our results with the N$^2$LO NN plus 3N interaction do not agree with the experimental binding energies for the upper $p$-shell, our results with the N$^4$LO$^+$ NN potential plus N$^2$LO 3NFs do agree with experiment throughout the $p$-shell within the combined numerical uncertainty estimates and the chiral truncation uncertainty estimates at the 68\% DoB. 

In future work we plan to extend these calculations to include consistent N$^3$LO 3NFs, which should bring the chiral truncation uncertainties down, and they may become comparable to the estimated numerical uncertainties.  
We therefore also intend to further reduce our numerical uncertainties; promising new developments include, among others, the use of Artificial Neural Networks~\cite{Negoita:2018kgi,Jiang:2019zkg,Knoll:2022abg} and Bayesian inference~\cite{Gazda:2022fte} for extrapolating NCSM binding energies to the complete basis.  The latter is particularly interesting, since with Bayesian methods for both the numerical and the chiral truncation uncertainties one can consider correlated uncertainties of different states.	This naturally leads to reduced uncertainties for excitation energies (compared to the uncertainties on the binding energies themselves), as well as e.g. neutron separation energies and	various	cluster	thresholds.

Last but not least, we plan to use the obtained wavefunctions, in combination with consistent $\chi$EFT operators, to evaluate other observables, in particular radii, charge densities, magnetic and quadrupole moments, and electroweak transitions. 

\section*{Acknowledgments}
We thank all members of the LENPIC collaboration for useful and inspiring discussions.
In particular, we would like to thank Evgeny Epelbaum, Dick Furnstahl, Kai Hebeler, Hermann Krebs and Jordan Melendez. 

\section*{Data Availability Statement}
The original contributions presented in the study are included in the article; further inquiries can be directed to the corresponding author.

\section*{Author Contributions}
All authors contributed to the calculations presented in this article, utilizing several different codes, developed at Iowa State University, the Forschungszentrum J\"ulich, and the Technische Universit\"at Darmstadt.  The initial draft for this article was written by PM, and all authors contributed to the subsequent discussions and final version of this article.

\section*{Funding}
This work was supported
by the US Department of Energy under Grants DE-FG02-87ER40371, DE-SC0018223 and DE-SC0023495, by the  Deutsche Forschungsgemeinschaft (DFG, German Research Foundation) Project-ID 279384907, SFB 1245, by DFG and NSFC through funds provided to the Sino-German CRC 110 “Symmetries and 329 the Emergence of Structure in QCD” (NSFC Grant No. 12070131001, Project-ID 196253076 - TRR 110), by the BMBF through Verbundprojekt 05P2021 (ErUM-FSP T07, Contract No. 05P21RDFNB), and by the MKW NRW under the funding code NW21-024-A.
This research used resources of the National Energy Research Scientific Computing Center (NERSC) and the Argonne Leadership Computing Facility (ALCF), which are US Department of Energy Office of Science user facilities, supported under Contracts No. DE-AC02-05CH11231 and No. DE-AC02-06CH11357, and computing resources provided  under the INCITE award `Nuclear Structure and Nuclear Reactions' from the US Department of Energy, Office of Advanced Scientific Computing Research. Further computing resources were provided on LICHTENBERG II at the TU Darmstadt and on JURECA and the JURECA Booster of the J\"ulich Supercomputing Center, J\"ulich, Germany. 

\section*{Conflict of Interest Statement}
The authors declare that the research was conducted in the absence of
any commercial or financial relationships that could be construed as a
potential conflict of interest.
 
\bibliography{paper_final}
%
\end{document}